\documentclass[reprint, amsmath,amssymb,prapplied, longbibliography, sd, superscriptaddress]{revtex4-2}

\usepackage{graphicx}
\usepackage{rotating}
\usepackage{amssymb}
\usepackage{amsmath}
\usepackage{mathptmx}
\usepackage{lipsum}
\usepackage{float}
\usepackage{siunitx}
\usepackage{hhline}
\usepackage{textcomp}

\begin{document}

\title{Superconducting microstrip losses at microwave and sub-mm wavelengths}
\author{S. H\"{a}hnle}
 \altaffiliation{s.haehnle@sron.nl}
\affiliation{SRON - Netherlands Institute for Space Research, Sorbonnelaan 2, 3584 CA Utrecht, The Netherlands}
\affiliation{Department of Microelectronics, Faculty of Electrical Engineering, Mathematics and Computer Science (EEMCS), Delft University of Technology, Mekelweg 4, 2628 CD Delft, The Netherlands}

\author{K. Kouwenhoven}
\affiliation{SRON - Netherlands Institute for Space Research, Sorbonnelaan 2, 3584 CA Utrecht, The Netherlands}
\affiliation{Department of Microelectronics, Faculty of Electrical Engineering, Mathematics and Computer Science (EEMCS), Delft University of Technology, Mekelweg 4, 2628 CD Delft, The Netherlands}

\author{B. Buijtendorp}
\affiliation{Department of Microelectronics, Faculty of Electrical Engineering, Mathematics and Computer Science (EEMCS), Delft University of Technology, Mekelweg 4, 2628 CD Delft, The Netherlands}

\author{A. Endo}
\affiliation{Department of Microelectronics, Faculty of Electrical Engineering, Mathematics and Computer Science (EEMCS), Delft University of Technology, Mekelweg 4, 2628 CD Delft, The Netherlands}
\affiliation{Kavli Institute of NanoScience, Faculty of Applied Sciences, Delft University of Technology, Lorentzweg 1, 2628 CJ Delft, The Netherlands.}

\author{K. Karatsu}
\affiliation{SRON - Netherlands Institute for Space Research, Sorbonnelaan 2, 3584 CA Utrecht, The Netherlands}
\affiliation{Department of Microelectronics, Faculty of Electrical Engineering, Mathematics and Computer Science (EEMCS), Delft University of Technology, Mekelweg 4, 2628 CD Delft, The Netherlands}

\author{D. J. Thoen}
\affiliation{Department of Microelectronics, Faculty of Electrical Engineering, Mathematics and Computer Science (EEMCS), Delft University of Technology, Mekelweg 4, 2628 CD Delft, The Netherlands}

\author{V. Murugesan}
\affiliation{SRON - Netherlands Institute for Space Research, Sorbonnelaan 2, 3584 CA Utrecht, The Netherlands}

\author{J. J. A. Baselmans}
\affiliation{SRON - Netherlands Institute for Space Research, Sorbonnelaan 2, 3584 CA Utrecht, The Netherlands}
\affiliation{Department of Microelectronics, Faculty of Electrical Engineering, Mathematics and Computer Science (EEMCS), Delft University of Technology, Mekelweg 4, 2628 CD Delft, The Netherlands}

\begin{abstract}
We present a lab-on-chip experiment to accurately measure losses of superconducting microstrip lines at microwave and sub-mm wavelengths. The microstrips are fabricated from NbTiN, which is deposited using reactive magnetron sputtering, and amorphous silicon which is deposited using plasma-enhanced chemical vapor deposition (PECVD). Sub-mm wave losses are measured using on-chip Fabry-P{\'e}rot resonators (FPR) operating around $\SI{350}{\giga\hertz}$. Microwave losses are measured using shunted half-wave resonators with an identical geometry and fabricated on the same chip.
We measure a loss tangent of the amorphous silicon at single-photon energies of $\tan\delta =3.7\pm0.5\times10^{-5}$ at $\sim\SI{6}{\giga\hertz}$ and $\tan\delta = 2.1\pm 0.1\times10^{-4}$ at $\SI{350}{\giga\hertz}$. These results represent very low losses for deposited dielectrics, but the sub-mm wave losses are significantly higher than the microwave losses, which cannot be understood using the standard two-level system loss model.
\end{abstract}

\maketitle
\section{Introduction}
Low-loss transmission lines are a fundamental requirement for integrated superconducting devices operating at both microwave and sub-mm wavelengths. 
At microwave frequencies, the primary driver for low-loss transmission lines is the development of qubits\cite{Barends2013QbitXmon, QBIT2019Supremacy} and microwave kinetic inductance detectors (MKIDs)\cite{Day2003MKIDs, Janssen13} based on high quality factor resonators\cite{Barends2010Loss, Megrant2012highQresonator}, as well as parametric amplifiers based on very long transmission lines $>\SI{100}{\lambda}$ \cite{ho_eom_wideband_2012}.
Additionally, high-impedance transmission lines can be used to further integrate microwave electronics onto the device chip \cite{Wagner2019HighImpedance, Colangelo2021HighImpedance}.
All these devices are predominantly realized using coplanar waveguide (CPW) technology, achieving losses corresponding to $Q_i>10^6$\cite{Barends2010Loss, Megrant2012highQresonator}. However, the planar nature of CPWs leads to large and complicated designs which can be difficult to scale. 
Multi-layer structures such as microstrips\cite{Mazin2010MicrostripMKID} and parallel-plate capacitors\cite{Beldi2019ParallelPlate} are preferable in order to create smaller devices and easily obtain high-impedance transmission lines, but these structures suffer from increased losses in the additional dielectric layer. 

At sub-mm wave frequencies, astronomical applications rely increasingly on integrated devices, such as multi-color/multi-polarization pixels \cite{Polarbear2008CMB_Multichroic}, phased array antennas \cite{Ade2015Bicep2} and on-chip filterbanks \cite{DeshimaNature,Superspec,Cataldo2018}. These usually use microstrips, as common mode excitation and radiation loss become serious issues at higher frequencies \cite{Haehnle2020,Spirito2013Commonmode}. However, these devices are usually based on Nb/SiO$_2$ technology, which has a $\SI{690}{\giga\hertz}$ cutoff due to the critical temperature of Nb and relatively high losses due to the SiO$_2$. 

The losses of microstrips at these frequencies and sub-Kelvin temperatures are generally attributed to the existence and excitation of two-level tunneling systems (TLS) in the bulk of the amorphous dielectric. While the macroscopic behaviour of TLS is relatively well understood, their microscopic origin is still for the most part unclear. Due to the lack of a microscopic understanding, development and investigation of low-loss dielectrics relies heavily on iterative cycles of deposition and measurement. These loss measurements are usually only carried out at microwave frequencies, under the assumption that material properties are comparable at sub-mm frequencies based on TLS theory \cite{Phillips1987TLSReview}. As a result, limited data is available at $f>\SI{100}{\giga\hertz}$ and $T<\SI{1}{\kelvin}$, where both signal generation and detection become increasingly challenging. 

Measurements comparing microwave and sub-mm wave loss were performed by Chang et al.\cite{Chang2015MicrostripLoss} and Gao et al.\cite{Gao2009MicrostripLoss} on Nb/SiO$_2$/Nb microstrips at $\SI{220}{\giga\hertz}$ and $\SI{110}{\giga\hertz}$ respectively, and by Endo et al. \cite{Endo2013MicrostripLoss} on NbTiN/SiN/NbTiN microstrips at $\SI{650}{\giga\hertz}$. These experiments combined quasi-optical techniques with different on-chip test devices exploiting either path length differences (Chang et al.) or resonant structures (Gao et al. and Endo et al.). However, these approaches suffer from large intrinsic uncertainties and are not sufficiently precise to study materials with a lower loss tangent. In the case of resonant structures, this can be mitigated by using long resonators at higher mode numbers, where coupler effects are suppressed.

In this paper, we present loss measurements of superconducting NbTiN/a-Si/NbTiN  microstrips at microwave and sub-mm wavelengths and at sub-Kelvin temperatures. We show TLS-like behaviour at microwaves, but find a significant increase in loss at sub-mm wavelengths which is inconsistent with the TLS standard model.

\section{Device Design}
\begin{figure*}[ht]
	\includegraphics[]{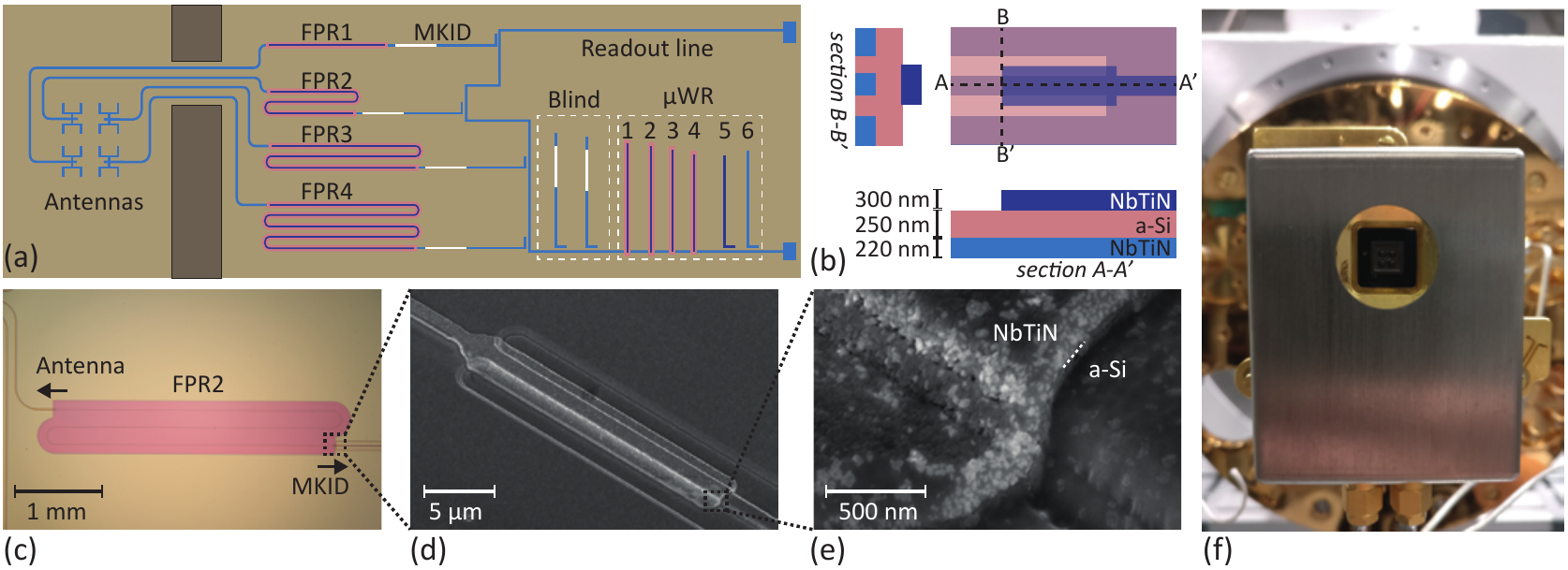}
	\caption{a) Chip schematic. NbTiN CPW structures are given in \textit{light blue}, the Al/NbTiN CPW of the hybrid MKIDs is given in \textit{white}, a-SI is shown in \textit{red} and the NbTiN microstrip lines are shown in \textit{dark blue}. One of the CPW microwave resonators (\textmu{}WR5) is also shown in \textit{dark blue} to highlight the use of the upper NbTiN layer for the CPW center line. b) FPR coupler schematic with two cross sections A-A' and B-B'. The schematic is shown top-down with semi-transparent layers, in order to clearly show the layout of the bottom NbTiN layer which is fully covered by a-Si. c) Optical microscope image of FP2. d) Angled scanning electron microscope (SEM) image of the FPR coupler. e) High-resolution SEM image of the microstrip open end. Overetch into the a-Si layer is visible, with the dotted white line indicating the border between NbTiN and a-Si. f) Picture of the device mounted in the cryostat. }
	\label{fig:device}
\end{figure*}

We have designed a chip which combines shunted microwave resonators (\textmu{}WR) and in-line sub-mm wave Fabry-P{\'e}rot resonators (FPR). The \textmu{}WR are coupled to a co-planar waveguide (CPW) transmission line (henceforth called readout line) which runs over the entire chip, terminating in bond pads connecting the chip to a sample holder via Al bond wires. The loss as a function of photon number can be obtained by measuring the readout line transmission around the resonance frequency of the shunted resonator and retrieving the loaded quality factor $Q_L$ and minimum transmission $S_{21,min}$ of the lorentzian dip. The internal loss factor $Q_i$ and coupling strength $Q_c$ then follow from the dip depth $S_{21,min}=Q_L/Q_i$ and the loaded quality factor, which can be expressed as
\begin{equation}
	\label{eq:QL_fun}
	Q_L(n) = \frac{nQ_{c,1}Q_i}{nQ_{c,1}+Q_i}
\end{equation}
where $n$ is the mode number of the resonance peak and $Q_{c,1}$ is the coupling strength at $n=1$, with $Q_c = nQ_{c,1}$ (see supplementary material).

For the sub-mm wave case, we excite the FPRs using a quasi-optical scheme, coupling radiation from a coherent source via an antenna and CPW line to the on-chip FPR. The relative power transmitted is measured as a function of source frequency using a microwave kinetic inductance detector (MKID) coupled to the far end of the FPR and read out using the same readout line as the \textmu{}WRs. This approach was previously used to measure CPW FPRs\cite{Haehnle2020}.

The design of the chip is given in Fig. \ref{fig:device}a). Six \textmu{}WRs with resonance frequencies around $\SI{6}{\giga\hertz}$ are coupled to the readout line. Of these, four are half-wave microstrip resonators using NbTiN/a-Si/NbTiN microstrips with a dielectric thickness of $h=\SI{250}{\nano\meter}$ and line width $w=\SI{2}{\micro\meter}$, and two are quarter-wave CPW resonators with a linewidth of $\SI{6}{\micro\meter}$ and slot width of $\SI{16}{\micro\meter}$. The CPW resonators are deposited directly on the Si substrate, with \textmu{}WR6 fully implemented in the lower NbTiN layer, while the center line of \textmu{}WR5 is implemented using the upper NbTiN layer. 

We implement four FPRs on the chip, where each FPR is connected to its own separate feeding network via two identical couplers, using the design shown in Fig. \ref{fig:device}b).  All four resonators are made with identical line and coupler geometry, but have different lengths $l_{FP}$ (FPR1: $\SI{5}{\milli\meter}$; FPR2: $\SI{10}{\milli\meter}$; FPR3: $\SI{20}{\milli\meter}$; FPR4: $\SI{50}{\milli\meter}$). This design was chosen to obtain resonators which, for the same frequency, have different mode numbers but identical $Q_i$ and $Q_{c,1}$.

We use a  CPW with $\SI{2}{\micro\meter}$ line and gap width between the antenna and the FPR, since we know its loss to be negligible at sub-mm wave frequencies \cite{Haehnle2020}. The MKIDs are quarter-wave resonators of a standard Al/NbTiN hybrid CPW design\cite{Janssen13}, where incoming power is absorbed in a short Al central line, leading to changes in the MKIDs microwave properties which are measured via the microwave readout line. 

\section{Device Fabrication}
We start the fabrication process with a $\SI{375}{\micro\meter}$ thick 4-inch Si wafer (dielectric constant $\epsilon_{r}=11.44$\cite{Lamb1996Materials}) coated on both sides with a $\SI{1}{\micro\meter}$ thick, low tensile stress ($\sim\SI{250}{\mega\pascal}$) SiN layer ($\epsilon_{r}=7$), deposited using low pressure chemical vapor deposition (LPCVD). 
The SiN on the device side is etched away almost everywhere, except for small patches below the MKID Al center lines \cite{Lorenza2018IEEE}. For the FPR and \textmu{}WR fabrication, we now start by depositing a $\SI{220}{\nano\meter}$ thick NbTiN layer ($T_c=\SI{15.1}{\kelvin}$, $\rho_n = \SI{138}{\micro\ohm\centi\meter}$) on the device side using reactive sputtering of a NbTi target in a nitrogen-argon atmosphere \cite{Thoen2016NbTiN}. This layer is patterned and etched to contain the microstrip ground plane and all CPW elements of the chip.
A $\SI{250}{\nano\meter}$ thick a-Si layer $\epsilon_{r}\approx 10$, deposited using plasma enhanced chemical vapor deposition (PECVD), serves as dielectric layer of the microstrip\cite{Bruno2020aSideposition}. 
We then define the microstrip lines in a second NbTiN layer of $\SI{300}{\nano\meter}$ ($T_c=\SI{15.0}{\kelvin}$, $\rho_n = \SI{104}{\micro\ohm\centi\meter}$), using the same process as the first NbTiN layer.
With the FPR and \textmu{}WR finalized, we finish fabrication of the MKIDs and microwave readout line using a $\SI{1}{\micro\meter}$ thick layer of polyimide LTC9505 and a $\SI{50}{\nano\meter}$ thick layer of Al ($T_c = \SI{1.25}{\kelvin}$) \cite{Haehnle2020Leaky}.
Finally, a $\SI{40}{\nano\meter}$ thick layer of $\beta$-phase Ta ($T_c=\SI{0.7}{\kelvin}$) is deposited on the backside and patterned into an absorbing mesh for stray light control\cite{Yates2017}.

As NbTiN and a-Si require the same dry etch agents, an overetch in the order of $\SI{40}{\nano\meter}$ is present for the lower layers, as highlighted in Fig. \ref{fig:device}e). The thickness of the lower NbTiN layer was increased accordingly to maintain designed antenna and MKID properties.

\section{Experimental Setup}
The \textmu{}WRs are characterized in a dark cryogenic setup, which is optimized for low-background MKID experiments as discussed in \cite{Janssen13,Pieter2014Nature}, and using a standard homodyne technique enabled by a commercial vector network analyzer (VNA). In order to achieve acceptable noise levels at the low microwave powers required to reach the single-photon regime, a $\SI{-36}{\decibel}$ attenuation on the input, and an amplifier with $\SI{36}{\decibel}$ gain on the output were added to the microwave readout chain at room temperature. 

For measurements of the FPR transmission, a $2\times2$ lens-array is mounted on the chip backside and aligned to the double-slot antennas. Each lens has a hyper-hemispherical shape with $\SI{2}{mm}$ diameter, creating a diffraction limited beam \cite{Filipovic1993DoubleSlotAntenna}. The chip is then placed in a Cu sample holder, which is surrounded by a tight-fitting mu-metal magnetic shield (see Fig. \ref{fig:device}f)). This assembly is mounted on the cold stage of a dilution refrigerator operated at $\SI{120}{\milli\kelvin}$. A TERABEAM 1550 (\textit{TOPTICA Photonics AG}) photomixer continuous wave (CW) source at room temperature emits a tunable signal in the range of $0.1...\SI{1}{\tera\hertz}$ with a step size of $\SI{10}{\mega\hertz}$ and a bandpass filterstack inside the cryostat that defines a frequency band centered at $f=\SI{346}{\giga\hertz}$. The source signal is attenuated with a beam splitter to keep the MKIDs in a linear operating regime. Frequency multiplexing readout electronics are used to enable simultaneous measurements of all FPRs\cite{SpacekidsMUX}. Details on the setup can be found in \cite{Haehnle2020}, which is identical except for the cryogenic unit.

\section{Results and Discussion}
\subsection{Microwave resonators}
\begin{figure}[ht]
	\includegraphics[]{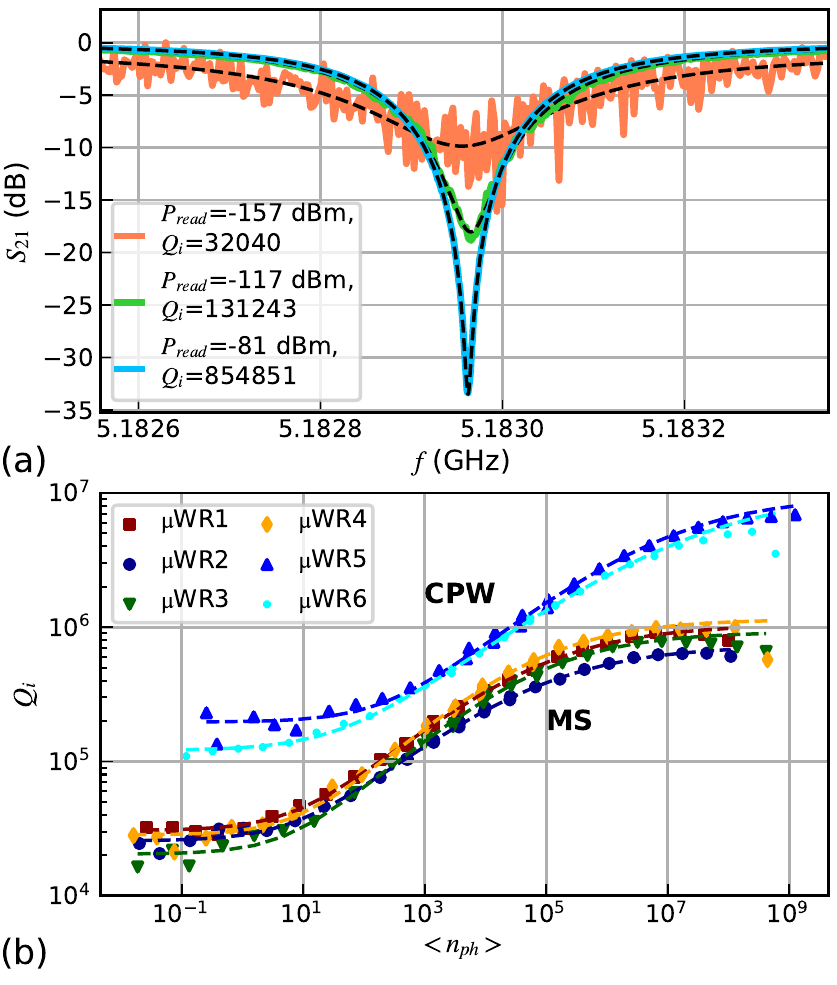}
	\caption{a) Transmission for \textmu WR1 at three different readout powers, with the respective Lorentzian fits shown by the black dotted lines. b) Measured internal quality factor $Q_i$ as function of $<n_{ph}>$ for all \textmu WRs shown as symbols, with fits of Eq. \ref{eq:TLS_strong} shown by dotted lines with corresponding colors.}
	\label{fig:results_ghz}
\end{figure}

We measure transmission dips of the \textmu{}WRs at $\SI{50}{\milli\kelvin}$ as a function of readout power in the range of $P_{read}=\SI{-65}{}...\SI{-157}{\decibel\meter}$, as shown exemplary in Fig. \ref{fig:results_ghz}a) for \textmu{}WR1. 
The internal quality factor $Q_i$ of the resonator is then determined at each power from a Lorentzian fit \cite{Khalil2012asymmetryFit}. We measure $S_{21}(f)$ from $\SI{4.7}{\giga\hertz}$ to $\SI{6.7}{\giga\hertz}$ in 4001 points at $T = \SI{60}{\milli\kelvin}$, excluding any points falling on a resonance feature, to create a $\SI{0}{\decibel}$  transmission reference.
Figure \ref{fig:results_ghz}b) shows the resulting $Q_{i}$ as a function of the average photon number in the resonator per $\lambda/2$, given by
\begin{equation}
	<n_{ph}> = \frac{P_{int}}{hf^2}
\end{equation}
where $P_{int}$ is the internal power, given for the \textmu{}WRs by $P_{int}=mQ^2P_{read}/\pi Q_c$ with $m=2$ or $m=1$ for  quarter-wave and half-wave resonators respectively. All \textmu{}WRs show a characteristic behaviour for TLS loss and can be fitted using \cite{molinaruiz2020origin}
\begin{equation}
	\label{eq:TLS_strong}
	\frac{1}{Q_i} = \frac{\tanh(\hbar\omega/2k_bT)}{Q_{i,0}(1+(<n_{ph}>/n_s))^{\beta/2}}+\frac{1}{Q_r}
\end{equation} 
where $Q_{i}$ has a minimum value $Q_{i,0}$ at photon numbers below the saturation value $n_s$, increases with a slope given by $\beta$ until other loss source dominate at high photon numbers, saturating at $Q_r$. 
As the CPW resonators do not contain a-Si, their TLS losses are dominated by the metal-substrate and substrate-air interfaces, with crystalline Si as substrate\cite{Wenner2011CPWloss}. Both $Q_{i,0}$ and $Q_r$ values (\textmu{}WR5: $Q_{i,0}\approx2.0\times10^5$, $Q_r\approx9.5\times10^6$. \textmu{}WR6: $Q_{i,0}\approx1.2\times10^5$, $Q_r\approx9.9\times10^6$) are comparable with the state of the art for CPW resonators made of sputtered NbTiN \cite{Barends2010Loss}, indicating an excellent film and interface quality for both the upper and lower NbTiN layers (sampled by \textmu{}WR5 and \textmu{}WR6 respectively) with no significant degradation due to the fabrication process. We find values of $\beta$ between $0.7$ and $0.8$ for all resonators, which is consistent with literature\cite{molinaruiz2020origin}. 

Losses in amorphous microstrips are generally dominated by TLS in the bulk dielectric layer\cite{Daalman2014TLS}. We assume that losses in the metal-substrate interface can be neglected here, due to the surface layer quality shown by the CPW resonators. Consequently, the internal loss factor of the a-Si film, can be obtained as $Q_{i,aSi}(\SI{6}{\giga\hertz})=27\pm4\times10^3$ from an average over all microstrip resonators and using
\begin{equation}
	\label{eq:FF_Qi}
	Q_{i,aSi} = FF\times Q_i
\end{equation}
with the filling factor $FF=0.96$ of the microstrip line, taking the a-Si layer overetch into account (see Fig. \ref{fig:device}e). This value corresponds to a loss tangent $\tan\delta = Q_i^{-1}=3.7\pm0.5\times10^{-5}$, which is expected for a-Si films \cite{Oconnell2008LossOverview}. We calculate the filling factor $FF$ by simulating a short microstrip line in CST and varying the loss tangent of the dielectric. The relation between simulated line loss and the set loss tangent is then given by $FF$.

\subsection{Fabry-P{\'e}rot resonators}
\begin{figure}[ht]
	\includegraphics[]{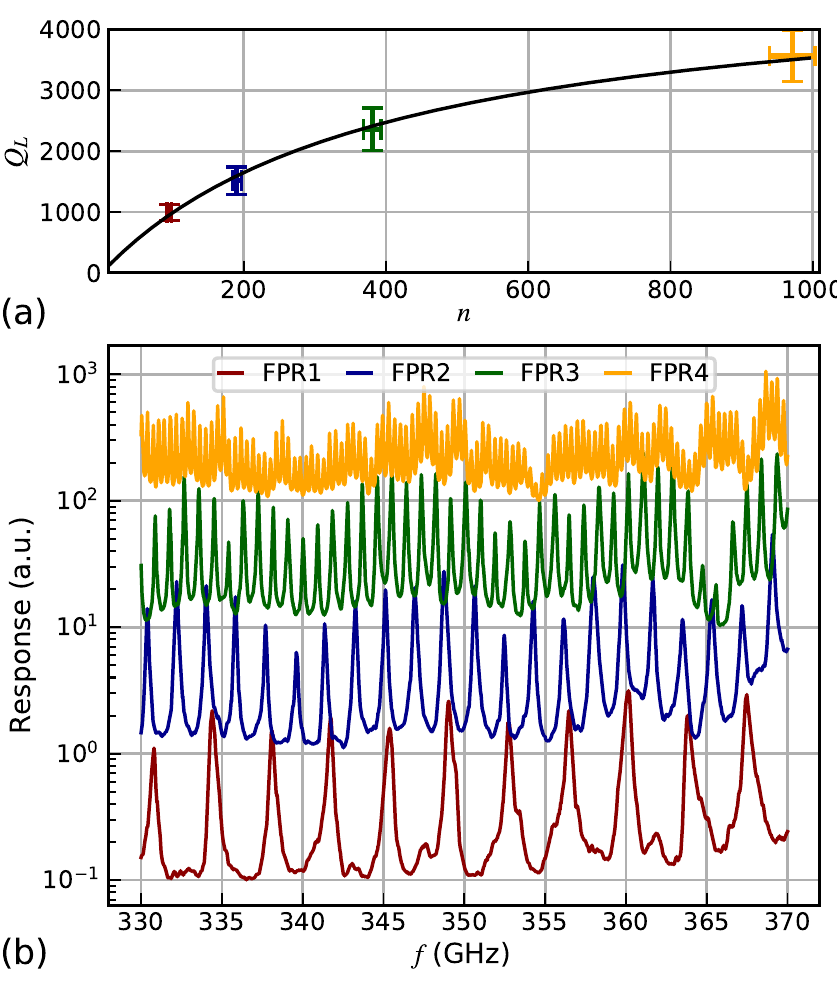}
	\caption{(a) Averaged loaded quality factor $Q_L$ as a function of average mode number in the four FPRs. The fit using Eq. \ref{eq:QL_fun} is shown by the black line. (b) Figure supporting figure a): Measured spectra of the FPRs, offset along the y-axis for better comparison. Line colors correspond to the markers in figure a).}
	\label{fig:results_thz}
\end{figure}

We measure the transmission through the four FPRs as a function of source frequency and apply a correction for directly coupled stray light, using the response of the blind MKID as introduced in \cite{Haehnle2020}. Figure \ref{fig:results_thz}b) shows the resulting spectra containing clear transmission peaks with an average frequency spacing $dF$ corresponding to the different resonator lengths. Variations in peak height, as well as secondary peaks visible in FPR1, can be explained qualitatively by standing waves before the FPR modifying the transmission through the first coupler with a wavelength corresponding roughly to the electrical distance between antenna and FPR coupler. For each resonance peak, we obtain $Q_L$ from a Lorentzian fit and the mode number from the resonance frequency $F_n$ using $n=F_n/dF$. The standing waves before the FPR also introduce an oscillation in the measured $Q_L$, due to a modified $Q_c$.

In this setup, we do not have an absolute reference for unity transmission and can therefore not use the height of the FPR peaks to obtain $Q_i$ and $Q_c$ independently. However, as we are measuring different resonators, each designed with different mode numbers $n$ at the same frequency, we can fit $Q_L(n)$ using Eq. \ref{eq:QL_fun} to retrieve $Q_{c,1}$ and $Q_i$. For this purpose, we take the mean value of $Q_L$ for each resonator, as plotted in Fig. \ref{fig:results_thz}a), while assuming that $Q_L$ is linear with frequency and that the standing wave oscillations are sampled well enough to be averaged out. The resulting mean value then corresponds to the value at the center frequency of the measurement band, $\SI{350}{\giga\hertz}$, while the error bars are a measure of both the statistical scatter and the linear slope over the averaged frequency range.
A fit to Eq. \ref{eq:QL_fun} then results in $Q_{c,1}=12.4\pm0.7$ and $Q_i=4.9\pm0.3\times10^3$.

Following Eq. \ref{eq:FF_Qi}, we then obtain the internal loss factor of a-Si at $\SI{350}{\giga\hertz}$ as $Q_{i,aSi}(\SI{350}{\giga\hertz})=4.7\pm0.3\times10^3$, corresponding to a loss tangent of $\tan\delta(\SI{350}{\giga\hertz}) = 2.1\pm 0.1\times10^{-4}$ which represents a significantly higher loss compared to the microwave regime, where $\tan\delta(\SI{6}{\giga\hertz}) =3.7\pm0.5\times10^{-5}$. Furthermore, we estimate the maximum incident power at the FPR to be less than $\SI{5}{\pico\watt}$, placing the FPR in the single-photon regime with $n_{ph}<8$.
The increased loss at sub-mm wave frequencies is in disagreement with the TLS standard model, where a constant loss is predicted in the single-photon regime when $\hbar\omega>>k_BT$, which is the case for both measurements.

A possible explanation of this discrepancy, could be a modified TLS model with a frequency dependent density of states $N(f)$. Such a model has been recently used to describe dephasing and noise due to TLS, combining interacting TLS with the power-law of $N(f)\propto(f^\mu)$ with $\mu\approx 0.3$ obtained from noise measurements\cite{Faoro2015TLSmodel}. We find that a value of $\mu\approx 0.4$ would resolve the discrepancy in our measurement. Note, that the power dependence of $\beta<1$ we find at microwave frequencies is also often identified with interacting TLS.
Alternatively, it is possible that we observe the low-frequency tail of resonant phonon absorption peaks at far-infrared wavelengths. Room temperature measurements of SiO$_x$ and SiN in the far-infrared regime indicate that such a low-frequency tail could have the relevant magnitude \cite{Cataldo2016SiliconOxide, Cataldo2012SiliconNitride}. 
Losses in the NbTiN film due to disorder effects should also be considered.  However, we consider this highly unlikely to be the dominant effect, since measurements on a NbTiN CPW line with a similar film yield a $Q_i\approx17\times10^3$ limited fully by radiation losses \cite{Haehnle2020}. We also find no indication in DC- or microwave-measurements that the NbTiN film quality of either the top- or bottom layer is degraded. The two CPW datasets in Fig. \ref{fig:results_ghz} represent a CPW line from each NbTiN layer.
Further experiments in the $100-1000\SI{}{\giga\hertz}$ frequency range, in combination with microwave measurements, are needed to clarify this issue.

\section{Conclusion}
In conclusion, we measured the losses of high-Q resonators made of NbTiN/a-Si/NbTiN microstrips in the single-photon regime at $\SI{6}{\giga\hertz}$ and $\SI{350}{\giga\hertz}$. Particularly, we demonstrate an effective method to independently measure $Q_i$ and $Q_c$ at sub-mm wave frequencies without an absolute power calibration, using multiple Fabry-P{\'e}rot resonators on a single chip. This method can be easily extended up to the bandgap frequency of NbTiN around $\SI{1}{\tera\hertz}$. We measure a very low loss tangent for the a-Si film at microwave frequencies, with a power dependence that is consistent with TLS theory. At sub-mm wave frequencies, we find an unexpected increase in the loss tangent, which requires further investigation to identify the root cause.

\appendix

\section{Mode Number Dependence of Qc}
In Goeppl et al. \cite{Goeppl08}, an equation for the coupling Q-factor of a FPR ($Q_{ext}=Q_c$) is derived as
\begin{equation}
	\label{eq:Qext}
	Q_{ext} = \frac{C}{2n\omega_0 R_L C_k^2}
\end{equation}
with the real load impedance $R_L=\SI{50}{\ohm}$, the coupling capacitance $C_k$ and the resonator capacitance $C=C_ll/2$ where $C_l$ is the capacitance per unit length and $l$ is the resonator length. This equation apparently shows a mode number dependence $Q_{ext}\propto1/n$, which is seemingly in conflict with the dependence $Q_c\propto n$ used in this paper, which is based on the equation
\begin{equation}
	\label{eq:Qc}
	Q_c = \frac{n\pi}{|S_{21,c}|^2}
\end{equation}
where $S_{21,c}$ is the transmission through a single coupler. In the following section we will show the derivation of Eq. \ref{eq:Qc}, the equivalence between Eq. \ref{eq:Qext} and Eq. \ref{eq:Qc}, the underlying reason for the apparent conflict in mode number dependence and finally why $Q_c\propto n$ should be used when considering the properties of a FPR.

Equation \ref{eq:Qc} is based on the definition of the quality factor
\begin{equation}
	\label{eq:Qdefinition}
	Q = \frac{\omega E_{stored}}{P_{loss}}
\end{equation}
where $E_{stored}$ is the total energy stored in the resonator, while $P_{loss}$ is the power loss per full resonator cycle. 

The power loss per cycle through a single coupler of the FPR is given by 
\begin{equation}
	\label{eq:ch2_Ploss_coupler}
	P_{loss} = NfE_{stored}|S_{2'1'}|^2
\end{equation}
where $S_{2'1'}$ is the transmission through the coupler and $N$ is the number of times the coupler is encountered per resonator cycle. For a half-wave resonator,  $N$ is given by the mode number as $N=1/n$. Combining Eq. \ref{eq:Qdefinition} and \ref{eq:ch2_Ploss_coupler} then results in
\begin{equation}
	\label{eq:ch2_Qc1_basic}
	Q_{c1} = \frac{n2\pi}{|S_{2'1'}|^2}
\end{equation}
The second coupler can then be included to obtain the total coupling Q-factor as 
\begin{equation}
	\label{eq:ch2_Qc_basic}
	\frac{1}{Q_{c}} = \frac{1}{Q_{c1}} + \frac{1}{Q_{c2}} = \frac{n\pi}{|S_{2'1'}|^2} 
\end{equation}
where the latter relation retrieves Eq. \ref{eq:Qc} for identical coupling strength on both sides ($Q_{c1}=Q_{c2}$).

In order to show the equivalence of $Q_{ext}=Q_c$, we start by taking the ABCD matrix of an isolated coupler as given by Goeppl et al.
\begin{equation}
	\left(
	\begin{matrix}
		A & B  \\
		C & D 
	\end{matrix}
	\right) = \left(
	\begin{matrix}
		1 & \frac{1}{i\omega_n C_k} \\
		0 & 1
	\end{matrix}
	\right)
\end{equation}
with $\omega_n=n\omega_0$ and retrieve $S_{21,c}$ from it as 
\begin{equation}
	S_{21,c} = \frac{2}{2+1/(i\omega_n C_kR_L)}
\end{equation}
where we use
\begin{equation}
	S_{21} = \frac{2}{A + B/R_L + CR_L+D}
\end{equation}
under the assumption that the characteristic line impedances $Z_0$ on either side of the coupler are matched to the load impedance $R_L$.

The absolute square $|S_{21,c}|^2$ then follows as
\begin{equation}
	\label{eq:s21coupler}
	|S_{21,c}|^2 = \frac{4\omega_n^2C_k^2R_L^2}{1+ 4\omega_n^2C_k^2R_L^2} \approx 4\omega_n^2C_k^2R_L^2
\end{equation}
where the approximation follows from the condition of a small capacitance ($\omega_nC_kR_L<<1$) as used by Goeppl et al. Note here the proportionality $|S_{21,c}|^2 \propto \omega_n^2$.

Combining Eqns. \ref{eq:Qext} and \ref{eq:s21coupler} then results in
\begin{equation}
	\label{eq:sub_one}
	Q_{ext} = \frac{2\omega_nC R_L}{|S_{21,c}|^2},
\end{equation}
which is not quite identical to Eq. \ref{eq:Qc}, but already follows its proportionality of $Q_{ext}\propto n$ due to $\omega_n=n\omega_0$.

We then use $v_{ph} = 1/\sqrt{(L_lC_l)}$, the line impedance $Z_0 = \sqrt{\frac{L_l}{C_l}}$ and $L_l$ the inductance per unit length to obtain the identity of 
\begin{equation}
	\frac{\omega_nC_l l}{2} = \frac{\omega_n l}{v_{ph}Z_0}
\end{equation}
which we can rewrite, using $\omega_n/v_{ph}=2\pi/\lambda$, $l/\lambda = n/2$ (for a half-wave resonator) and the matched impedance condition $Z_0=R_L$, as
\begin{equation}
	\label{eq:omegacl}
	\frac{\omega_nC_l l}{2}= \frac{n\pi}{R_L}.
\end{equation}

Substituting Eq. \ref{eq:omegacl} in \ref{eq:sub_one} we finally obtain the form of Eq. \ref{eq:Qc}
\begin{equation}
	Q_{ext} = \frac{n\pi}{|S_{21,c}|^2} = Q_c.
\end{equation}
Note here, that the matched impedance condition was only used to easily show the equivalence, but is not necessary for either equation to be used on its own. For Eq. \ref{eq:Qext} it is explicitly included due to the presence of $R_L$, while Eq. \ref{eq:Qc} includes it implicitly as $S_{21,c}$ is obtained from EM-simulations and can therefore account for any impedance mismatch. Additionally, any impedance mismatch does not factor into the mode number dependence.

In order to explain the apparently conflict in mode number dependence, we suggest, that while Eq. \ref{eq:Qext} is based on a flawless derivation and provides accurate fits to measurements\cite{Goeppl08}, it conflates mode number dependence with frequency dependence of the coupler and therefore does not provide an accurate physical interpretation of the mode number dependence.

The frequency dependence of the coupler can be understood from Eqns. \ref{eq:s21coupler} and \ref{eq:sub_one}: The couplers used in both of the papers considered here are fundamentally high-pass filters operated in the stop-band regime and their transmission is therefore proportional as $|S_{21,c}|^2 \propto \omega^2$. Note, that we omit the subscript $n$ here, as this proportionality exists independent from the resonant behaviour of the FPR and must therefore be considered separate from the mode number.  

Goeppl et al. argue, that Eq. \ref{eq:Qext} shows $Q_{ext}\propto 1/n$. However, its primary dependence should be understood as $Q_{ext}\propto 1/\omega_n$ which, when taking the coupler transmission into account, becomes $Q_{ext}\propto \omega_n/\omega^2\propto n/\omega^2$ (see Eq. \ref{eq:sub_one}).
In contrast, Eq. \ref{eq:Qc} shows a clear separation of coupler properties ($S_{21,c}$) and mode number. 

The difference shown here is usually not significant for single resonators, as both equations will yield the same results for $Q_c$. However, it needs to be taken into consideration when one wants to understand and exploit the fundamental behaviour of a transmission line FPR.

\bibliography{bibliography}

\end{document}